\documentstyle[12pt]{article}
\def\beq{\begin{equation}}
\def\eeq{\end{equation}}
\def\nn{\nonumber}
\def\mev{\,{\rm MeV}}
\def\gev{\, {\rm GeV}}
\def\csb{\Lambda_{CSB}}
\def\qcd{\Lambda_{QCD}}

\def\ttr{{\rm Tr}}
 \def\ih#1{{i#1\over2}}
\def\ch{\bar{H}_a^i}
\def\cha{\bar{H}^i}
\def\hi{H_i}
\def\hj{H_j}
\def\hai{H^a_i}
\def\hbi{H^b_i}

\def\hbarq{\overline{h}}
\def\hq{{h}}

\def\vmu{v_\mu} \def\dmu{\partial^\mu} \def\xidag{\xi^\dagger}
 
\def\Oone{{{\cal O}_1}}
\def\Otwo{{{\cal O}_2}}
\begin{document}
\title{The QCD Scale in the Heavy Quark Expansion
\thanks{This work is supported
in part by funds provided by the U.S.
Department of Energy (DOE) under contract \#DE-AC02-76ER03069 and in
part by the Texas National Research Laboratory Commission
under grant \#RGFY92C6.\hfill\break
\vskip 0.05cm
\noindent $^{\dagger}$National
Science Foundation Young Investigator Award.\hfill\break
Alfred P.~Sloan
Foundation Research Fellowship.\hfill\break
Department of Energy Outstanding Junior
Investigator Award.\hfill\break
CTP\#2167\hfill November 1992}}
\renewcommand{\baselinestretch}{1.0}
\author{Lisa Randall$^{\dagger}$ and Eric Sather\\
Massachusetts Institute of Technology\\
Cambridge, MA 02139\\
}
\date{}
\maketitle
\vskip-5in
November 1992 \hfill  MIT-CTP\#2167
\vskip5in
\renewcommand{\baselinestretch}{1.2}
\abstract{
We argue that consistency of the combined heavy quark  and
 chiral effective lagrangian requires  the QCD scale
which multiplies $1/M$ in the heavy quark expansion to be
the chiral symmetry breaking scale,  $\Lambda_{CSB}$,
rather than  the QCD scale, $\Lambda_{QCD}$. This means
that either there is large uncertainty in the
accuracy with which the heavy quark effective theory
can be applied to $c$ quarks or  the cutoff scale
of the heavy quark chiral effective  theory is lower than has
been assumed.
\thispagestyle{empty}
\newpage

\section{Introduction}

The heavy quark theory  \cite{hq} tells us that in the limit
that quarks are infinitely massive, symmetries exist
between quarks of different flavor and spin.  Of course,
in the real world, quarks are not infinitely massive,
so finite mass corrections must be taken into account.
This means that there are corrections
to the symmetry relations  (aside from calculable QCD
corrections)  which are suppressed by
factors of $1/M$, where $M$ is the mass of the heavy quark.
There must be a  dimensionful factor to compensate
this mass suppression.  Although we don't know precisely the
scale, we know that  it is some
QCD scale, which the heavy quark expansion takes to be of
order $\Lambda_{QCD}$, about a few hundred MeV \cite{qcd}.

Recent work \cite{wise} has incoporated low energy chiral dynamics
into a heavy quark theory.  In this theory, one can
consider the interactions of low energy pions and kaons with
the heavy mesons.  In this note, we argue that consistency
of the chiral and heavy quark expansions requires that the
QCD scale which multiplies $1/M$ in the heavy quark expansion
    is the chiral symmetry breaking
    scale,  $\Lambda_{CSB}$, rather than  $\Lambda_{QCD}$.

    \section{The QCD Scale}

We present two arguments that the scale appearing in the heavy
quark expansion should be $\csb$ and not $\qcd$.  The first
argument is a straightforward extension of the usual
one--loop argument
about the chiral symmetry breaking scale.
That is, it is an application of naive dimensional analysis (NDA)
\cite{nda} to the heavy quark chiral theory.
The second argument
is based on the assumption that
in the chiral lagrangian there is a consistent expansion
 in flavor symmetry breaking; that is,
effects depending on the current quark mass, $m$, of a light quark,
 should be suppressed by $m/\csb$ relative
to the leading terms.

Consider as an example the heavy quark operator
\beq
{\cal O}^h=\hbarq^i {D^2 \over 2 M_i} \hq^i,
\eeq
where $i$ is  the flavor of the heavy quark.
This operator will match onto $1/M$ suppressed terms in
the effective theory of the heavy meson, $\hi$.  One of these terms
will be
\beq
\Oone^H=\ttr \cha  {\partial^2 \over 2 M_i} \hi.
\eeq
Another will be the tree level
term
\beq
\Otwo^H={\Lambda^2 \over 2 M_i} \ttr \cha \hi.
\eeq
By consistency, we can estimate what $\Lambda$ should be.

We assume the by now standard
heavy meson
effective lagrangian, given by
\begin{eqnarray}
{\cal L}&=&\sum_i  \left\{ -i\ttr[{\ch\vmu\dmu\hai}]
   +\ih{ }\ttr[{\ch\hbi}]\vmu              (\xidag\dmu\xi+\xi\dmu\xidag)^a_b
\nn\right.\\
 &+&\left.
 \ih{g}\ttr[{\ch\hbi\gamma_\mu\gamma_5}](\xidag\dmu\xi-\xi\dmu\xidag)^a_b
     \nn \right \},\\
\end{eqnarray}
where we have summed over heavy quark flavors.  Here $H_a^i$ is the
heavy meson field with heavy quark index $i$ and light quark index $a$,
$v$ is its velocity, and $\xi=\exp{(i \pi^\alpha T^\alpha/f_\pi)}$.

Now consider mass suppressed corrections to  the heavy meson current,
given by
\beq
 L^{\nu \, i}_j= \ttr  \cha \gamma^\nu (1-\gamma_5) \hj.
 \eeq
At tree level, the matrix element of the current will also contain
mass suppressed terms, proportional to $\Lambda'/M$. We estimate
what we expect the loop calculation to contribute to such a mass suppressed
operator.
 We estimate the contribution from the diagram in which  there
 is one insertion of the current, $L^{\nu\, i}_j$ and one insertion of
  $\Oone^H$
 on the heavy meson line and a pion is emitted and absorbed
 through the axial coupling proportional to $g$.
 Because we are interested in determining the scale
 which multiplies $1/M$, we reason
 as was done previously  \cite{georgi,nda} and assume a cutoff regulator.
 Recall that the reasoning there was to estimate the one--loop contribution
 to a given operator, and to choose the chiral symmetry breaking scale
 in such a way that loop renormalization of a
 counterterm did not exceed the tree value.
 The reasoning here is slightly different, because we assume
 we know that the cutoff scale of chiral dynamics is $\csb\approx 4 \pi f_\pi$,
 and from this we wish to determine the tree term in the lagrangian.
 The assumption is however
 the same; we do not want a loop contribution
 to exceed the tree level coefficient.
  A naive estimate
 of the loop yields a correction to
 the original current with coefficient of order $\csb^3 /(16 \pi^2 f_\pi^2 M)$.
 Here, the $1/(16 \pi^2)$ comes from the loop, the factor
  of $1/M$ from the insertion of the mass suppressed operator,
  and the $1/f_\pi^2$ from the pion couplings. To get the right
  dimensions requires three factors of the cutoff in the numerator.
  In order that the loop amplitude does not exceed the tree
  amplitude significantly, we must have $\Lambda'=\csb$, that
  is the contribution to the matrix elements of mass suppressed
  operators is determined by the chiral symmetry breaking scale, $\csb$.

  Similar reasoning would imply  that $\Lambda$ in eqn. (3) is
  also $\csb$. This would follow from the estimate of a one--loop
  diagram with $\Oone^H$ inserted.
 The same dimensional factor $\csb$ was
 used by Georgi in a recent paper \cite{ddbar}, where he employed NDA.

  For those who are unhappy with estimates based on a cutoff regulator,
 we present an alternative argument based on the assumption
 that a consistent chiral expansion incorporates SU(3) symmetric
 operators, with symmetry breaking operators suppressed by
 explicit chiral symmetry breaking factors. One of these factors
 is the quark mass matrix $m_q$, which is a diagonal matrix
 proportional to  the light quark masses, $m_u$, $m_d$, and $m_s$.
 From the usual chiral lagrangian expansion, we know that these
 dimensionful symmetry breaking parameters must occur suppressed
 by the chiral symmetry breaking scale, $\csb$, relative to the
 leading term. Using this fact, we once again show
 that the scale $\Lambda$ must be taken as $\csb$ for consistency.

 This sort of calculation was considered in ref. \cite{rs}, where a
 large radiative correction to the $1/M$ suppressed
 chromomagnetic operator was generated at one loop.
 Although it is not necessary, the argument here is simplest
 if we insert the operator $\Oone^H$ not once but twice, to
 generate a correction to the current at order $1/M^2$. Using
 dimensional regularization, one would generate a correction
 to the current proportional to the relevant mass scale internal
 to the loop raised to the appropriate power. To be specific, let's consider
 the kaon loop contribution, which will be proportional
 to $m_K^4/(\csb^2 M_c^2) \approx m_s^2 /M_c^2$. We rewrite this
 to make the chiral expansion parameter explicit as
 $(m_s/\csb)^2 (\csb/M_c)^2$. Since this is the chirally
 suppressed term, as argued above,
 the lagrangian  should contain a leading order
 term, not suppressed by SU(3) breaking,
 of order $(\csb/M_c)^2$.
 In dimensional regularization, this term is not manifest in
 the loop calculation, because it arises from matching the full
 to the effective theory. However,
 the existence of a lagrangian
 which is SU(3) symmetric at leading order and  in which current quark
 mass contributions are suppressed by $m_s/\csb$ relative to the
 leading order terms
 means that the term proportional
 only to $(\csb/M_c)^2$ must be there for consistency.
  This gives us the same conclusion as
 the previous argument. The only way to prevent large renormalizatons
 of the tree level operators is to assume that the relevant
 mass suppression factor is $\csb/M$.

 Although we have illustrated our point with specific operators,
 the general argument should be clear. Because we have the
 same parameters appearing in the chiral lagrangian with and
 without the heavy mesons, it is inconsistent to choose the
 scales differently in the case that the heavy meson is and is not
 present. At the diagrammatic level, the same momentum runs
 through the heavy meson line
 and the pseudogoldstone boson lines.
The effective theory incorporates the scale of the dynamics
of the light degrees of freedom, or the momentum with which they
can recoil, in the cutoff.
 The QCD degrees of freedom in the heavy meson are
 allowed to have momentum up to the cutoff of the effective
 theory.
 The largest energy of the virtual degrees of freedom characterizes
 the expansion.
  If the derivative expansion of the heavy quark chiral effective theory
 is indeed applicable up to pion and kaon momenta of the order of the
 chiral symmetry breaking scale, then consistency demands
 that this is also the scale
 which appears in the matrix elements of $1/M$ suppressed operators,
 since it is the scale of momenta in the loops.
 This is
 true independently of whether the original
 operator involved derivatives or the gluon field.  All the scales
 should be consistently determined by naive dimensional analysis \cite{nda}.

  It should be emphasized that this is not just a statement about the
heavy quark chiral effective theory, although
it assumes the {\it existence} of this theory,
and that it is valid up to the scale $\csb$.
 With this assumption, once we have shown that the scale of the
 heavy meson lagrangian is $\csb$, then the  expansion
 in  inverse heavy quark mass
 requires that this is also the scale which characterizes
 matrix elements
 of operators
 in the heavy quark theory between the physical meson states.
 If the heavy quark expansion
is valid, matrix elements of quark operators suppressed by any given power
of $1/M$ must match onto heavy meson operators suppressed by the
same power of $1/M$.  Then, for the matrix elements to agree,
which is the requirement of the heavy meson lagrangian,
both theories must have the $M$ suppression compensated by the same scale,
which we have just demonstrated is $\csb$ in the heavy meson lagrangian.
This is therefore the scale which characterizes matrix elements
in the heavy quark theory as well.   For example, the matrix
elements of the heavy quark operator ${\cal O}^h$ between meson states
should be approximately equal to that of $\Otwo^H$, implying
the relevant scale in the heavy quark theory is also $\csb$.

It is of course possible that chiral expansion about the heavy meson
does not exist, or that  the heavy meson chiral
effective theory is not valid up to $\csb$.
  There might be a smaller
 cutoff beyond which the derivative expansion breaks down.
 It is then this lower cutoff which would set the scale
 for the $M$ suppressed matrix elements, so
 that they  are small.
 However, if this scale is indeed significantly lower than $\csb$,
 the heavy meson chiral effective theory would not be of much use,
 as the kaon mass would be comparable to the cutoff, and even
 pion loops might not be reliable.

It has been argued \cite{cdg} that the relevant scale for the cutoff
 is always some physical degree of freedom, which for the standard chiral
 lagrangian might be the $\rho$ mass.  In the case of the heavy
 meson lagrangian, this mass might be that of an excited state
 which has not been included. This state could occur as a virtual
 intermediate state; the mass difference between this state
 and the low lying heavy meson states might be what sets the
 scale of the cutoff, rather than $4 \pi f_\pi$.  If this
 is true, it would indicate that the cutoff of the heavy meson
 chiral lagrangian could be considerably lower than $\csb$. For example,
 the as-yet unobserved
 states with the total angular momentum of the light degrees of
  freedom $j=1/2$ and with orbital angular momentum $l=1$ have been
  predicted to be split from the low lying meson states by
  only $500 \mev$  \cite{gi}.

 We conclude that it is not possible to consistently treat
 the matrix elements of heavy quark currents as an expansion in
 $\qcd/M$ and the heavy meson lagrangian as a theory whose cutoff
 is $\csb$.  If indeed the chiral lagrangian is valid up to the
 chiral symmetry breaking scale, this also sets the scale for
 the mass suppressed matrix elements. Alternatively, the
 cutoff for the validity of the heavy quark chiral effective theory
 could be a much lower scale. This would however cast doubt
 on the utility of the heavy quark chiral lagrangian.

 \section{Discussion and Conclusions}

 The implications of this result are unclear. It was already
 known that there are potential problems with treating a $c$ quark
 as heavy.  If there are various small factors which go the right
 way, higher order terms in $m_c$ might be as small as desired
  by practitioners of the heavy quark theory.  However, it
 is clear that at least formally, consistency of the heavy quark and
 chiral expansions requires that heavy quark operators be
 suppressed by $\csb/M$.

 It is interesting to note that sum rule calculations have indeed
 yielded some large values for $\Lambda$ for particular operators.
 For example, the matrix element of $\Oone$
 between heavy mesons was defined in ref. \cite{n} as $2 M \lambda_1$,
 where QCD sum rules \cite{n} gave $\lambda_1 \approx 1 {\rm GeV}^2$. Other
 estimates have been done \cite{fn} based on a constituent quark model.
 Since the mass of the constituent quark lies squarely between the QCD
 and chiral symmetry breaking scale, it can be said to fit
 with either assumption for $\Lambda$. The spin splitting of the
 mesons indicates a QCD scale which again lies in between $\qcd$ and $\csb$.

 Probably the most important reason for a better determination
 of $\Lambda$ is to determine how accurately we can hope to
 extract KM angles using the heavy quark effective theory.
 Falk and Neubert \cite{fn} did a detailed analysis based
 on a constituent quark model and QCD sum rule estimates of the
 various matrix elements, where they find 1 to 3~\% corrections
 from mass suppressed operators.
 Our results would indicate that comparable errors are incurred
 just by neglecting the SU(3) {\it violating} contributions
 to the matrix elements proportional to the strange quark mass.
 If these are indeed suppressed corrections to leading order operators,
 the error in the extraction of $V_{cb}$ could be substantially
 larger than a few percent.

 If it is indeed true that the matrix element of the mass suppressed
 operator is set by the chiral symmetry breaking scale, it would
 appear that the heavy quark expansion is useless for an extraction
 of $V_{cb}$. However, if we are lucky, this might not be the case.
 In the particular model for the higher order coefficients employed
 by Falk and Neubert,
 there were fortuitous cancellations which made the effects
 of higher order terms small.
 However, their work also indicates
 some universal factors which make it conceivable that one could
 hope to extract the KM angle at the $10-20 \%$ level.  First is
 the fact that the heavy quark expansion is really
 an expansion in $1/2M$. Of course this factor of 2 can be compensated
 by 2's in chiral coefficients, but if we are lucky, the coefficients
 might all be less than unity (after dimensional analysis factors
 of order of 1 GeV have been extracted). In this case, $1/M^2$ corrections
 might be suppressed by $(1\gev/2 M_c)^2 \approx 10 \%$.  Even for $B \to D$,
 which is not protected by Luke's theorem \cite{luke} so that $1/M$ corrections
 are present, there is the Voloshin-Shifman factor \cite{vs},
  $S=(m_B-m_D)^2/(m_B+m_D)^2 \approx 0.23$,  multiplying the leading $1/M$
 corrections. Therefore, these
 corrections might also be at the level of  10 \%.

 The considerations of this paper also apply to the baryon lagrangian.
 This would severely compromise the utility of the baryon lagrangian.
 Here, the expansion parameter would be expected to be $\csb/2M_{\rm nucleon}$,
 which is not small.

 It might well be that the mass suppressed
 matrix elements are smaller than suggested
 in this note.
 Alternatively, the chiral lagrangian for heavy mesons might
 not be valid, or
 fail at a scale considerably lower than the standard chiral
 symmetry breaking scale. However, it is clear that until there is
 a firm argument that the estimate based on chiral loop estimates
 is incorrect, the uncertainty in extracting KM angles, even if it occurs
 only at $1/M^2$, is very large.
 It would
 be very useful to have measurements of mass suppressed heavy
 quark matrix elements to determine whether mass suppressed
 operators can be as large as implied by these estimates.

  \section*{Acknowledgements}
  We thank Adam Falk  for useful discussions.
  This work is supported in part by funds provided by the U. S. Department
  of Energy (D.O.E.) under contract \#DE--AC02--76ERO3069.


\begin{thebibliography}{9}
  \bibitem{hq}See H. Georgi (1991) {\sl Lectures presented at the Theoretical
  Advanced Study Institute, Boulder (World Scientific)}, to be published;
  M. Wise (1991){\sl Lectures presented at the Lake Louise Winter Institute,
  Caltech preprint} CALT-68-1721, B. Grinstein (1991) {\sl Proc. High Energy
  Phenomenology Workshop, Mexico City, eds. R. Huerta and M. Perez,
  SSCL preprint } 91-17, and refs. therein.
  \bibitem{qcd} N. Isgur and M. Wise, Phys. Lett. {\bf B 237} (1990) 527;
  A. Falk, B. ~Grinstein, and M. Luke, Nucl. Phys. {\bf B 357} (1991) 185.
   \bibitem{wise} M. Wise, Phys. Rev. {\bf D45} (1992) 3021.
  \bibitem{nda} A. Manohar and H. Georgi, Nucl. Phys. {\bf B 234} (1984) 189;
  H. ~Georgi and L. Randall, Nucl. Phys. {\bf B 276} (1986) 241.
  \bibitem{georgi} H. Georgi, ``Weak Interactions and Modern Particle
  Theory", Benjamin/Cummings Publishing Co., Inc. (1984).
  \bibitem{ddbar} H. Georgi HUTP-92/A049
  \bibitem{cdg} R. S. Chivukula, M. Dugan, and M. Golden , BUHEP-92-18.
  \bibitem{gi} S. Godfrey and N. Isgur, Phys. Rev. {\bf D32} (1985) 189.
  \bibitem{rs} L. Randall and E. Sather, MIT-CTP-2166
  \bibitem{fn} A. Falk and M. Neubert, SLAC-PUB-5897, 1992.
  \bibitem{n} M. Neubert, Phys. Rev. {\bf D45}, 2451 (1992).
  \bibitem{vs} M. Voloshin and M. Shifman, Yad. Fiz. {\bf 45}, 463 (1987)
  [Sov. J. Nucl. Phys. {\bf 45}, 292 (1987); {\bf 47}, 801 (1988) [{\bf 47},
  511 (1988)].
  \bibitem{luke} M. Luke, Phys. Lett. {\bf B 349}, 598 (1991).
  \end{thebibliography}
\end{document}